\documentclass[12pt,preprint]{aastex}

\shorttitle{CO 2-1 Emission in NGC~4258}
\shortauthors{Sawada-Satoh et al.}

\begin{document}


\title{Structure and Kinematics of CO (J=2-1) Emission in the Central 
Region of NGC~4258}


\author{S. Sawada-Satoh \altaffilmark{1} P. T. P. Ho \altaffilmark{1,2}
S. Muller \altaffilmark{1} S. Matsushita \altaffilmark{1} J. Lim\altaffilmark{1}}





\altaffiltext{1}{Academia Sinica Institute of Astronomy and Astrophysics, 
P.O. Box 23-141, 
Taipei 10617, Taiwan}
\altaffiltext{2}{Harvard-Smithsonian Center for Astronphysics, 60 Garden Street, Cambridge, MA 02138, USA}


\begin{abstract}
We present $^{12}$CO ($J=$2-1) observations towards  
the central region of the Seyfert 2 galaxy NGC~4258 
with the Submillimeter Array (SMA).  
Our interferometric maps show 
two arm-like elongated components 
along the major axis of the galaxy, 
with no strong nuclear concentration. 
The CO (2-1) morphology and kinematics are 
similar to previous CO (1-0) results. 
The velocity field of the components agrees 
with the general galactic rotation, 
except for the east elongated component, 
which shows a significant velocity gradient 
along the east-west direction. 
In order to account for the velocity field, 
we propose the kinematical  model 
where the warped rotating disk is also expanding. 
The line ratio of CO(2-1)/CO(1-0) reveals that 
the eastern component with the anomalous velocity gradient appears 
to be warmer and denser. 
This is consistent with the gas in this component being closer to the center,  
being heated by the central activities, and possibly interacted by 
expanding motions from the nuclear region. 
\end{abstract}





\section{Introduction}

Molecular gas in the central regions of Seyfert galaxies provides important
clues to the nuclear environment of the Active Galactic Nuclei (AGN).
It is generally proposed that the activity of the AGN is driven 
by the accretion of gas
onto a massive black hole in the center of the host galaxy.
The actual structure and dynamics of the nuclear gas will 
provide some constraints on the accretion process. 
There have been many molecular line studies of the  
central regions of galaxies with interferometry. 
Recent molecular gas imaging surveys in Seyfert galaxies 
show that molecular clouds in circumnuclear region have diverse 
morphologies (e.g. Kohno et al. 2001, Garc\'{i}a-Burillo et al. 2003). 
Different CO lines are sometimes found to exhibit 
different distributions in the same galaxy.
For example, in the case of M51,
a Type-2 Seyfert galaxy,
the CO (1-0) line emission is mainly associated with 
galactic arms and 
a weak nuclear peak (Sakamoto et al. 1999).
However, in the CO (3-2) line, the intensity in temperature 
of the central 
concentration is higher than in the CO (1-0) line (Matsushita et al. 2004). 
 This suggests that the nuclear regions may have warm  
and dense gas which is better sampled with higher excitation lines. 
Currently, M51 is the only Seyfert galaxy 
where several CO lines have been well studied with 
millimeter and submillimeter interferometers,
and it is important to know whether the presence of 
warm gas in the nuclear regions is in fact common for other Seyferts.

NGC~4258 is a nearby spiral galaxy whose nuclear activity 
is classified as Seyfert 2 or LINER 
(Heckman 1980; Stauffer 1982; Makishima et al. 1994; Ho et al. 1997), 
and whose distance is 7.2~Mpc (Herrnstein et al. 1999).
This galaxy is also known for bisymmetric ``anomalous arms'' 
in H${\alpha}$, radio continuum and X-ray emission that
span nearly the full visual extent of the galaxy 
(e.g. van der Kruit et al. 1972, Pietsh et al. 1994, Wilson et al. 2001).
The anomalous arms have been interpreted as jets from its galactic nucleus 
(Ford et al. 1986; Martin et al. 1989; Cecil et al. 1992, 1995, 2000; 
Dutil et al. 1995; Daigle and Roy 2001) or 
as gas flow due to a bar (Cox and Downes 1996). 

NGC~4258 is also well known to exhibit a sub-pc-scale molecular disk
(P.A.$=86\pm2^{\circ}$, Inclination$=83\pm4^{\circ}$) 
as traced by H$_2$O maser emission 
(Miyoshi et al. 1995, Herrnstein et al. 1999).   
The major axis of the molecular disk is along the east-west direction 
although the P.A. of the host galaxy is around $-30^\circ$ 
(e.g. van Albada 1980, Cox and Downes 1996), 
and the angle between the two rotation axes of the sub-pc-scale disk 
and the galactic disk is $\sim120^{\circ}$ 
(Miyoshi et al. 1995). 
This indicates a significant misalignment between the axes of rotation for the entire galaxy and for the nuclear disk.
Until now, the morphological and kinematical connections 
between the host galaxy and the nuclear disk have not been elucidated.

In NGC~4258, the CO (1-0) and (2-1) lines have been detected 
in single dish observations with a large velocity range 
from $\sim$200 to $\sim$700 km~s$^{-1}$ 
and are concentrated to the anomalous arms in the nuclear region   
(Sofue et al. 1989, Krause et al. 1990, Cox and Downes 1996).
Interferometric studies of the CO (1-0) line emission have been done 
with OVRO (Plante et al. 1991), 
BIMA  (Helfer et al. 2003) and PdBI (Krause et al. 2005),  
and they show  two parallel elongated structures 
along a direction of P.A. $\sim150^{\circ}$ associated with the central part of  the anomalous arms, 
although the peak position of  H$\alpha$ is in a depression 
between the two CO structures   
(Martin et al. 1989; Plante et al. 1991; Cecil et al. 1992; Krause et al. 2005). 
To interpret the CO distribution, three models had been proposed.  
One model suggest that jets emanate from the AGN, 
and dig tunnels into the molecular gas. 
The jets then separate a big molecular cloud into two regions 
and create the depression between the two regions.  
Therefore, CO emission may be tracing the molecular gas 
which has been interacted with the jets  
(Martin et al. 1989, Plante et al. 1991). 
Plante et al. (1991) developed the model 
which superposes uniform expansion 
away from the nucleus 
in addition to a velocity pattern which is similar to that of the solid body rotation.  
Still another model suggests that the CO molecular gas is tracing an elongated bar structure, 
and that  the ``anomalous arms" may be region of shocked gas due to the 
presence of a bar.  
Cox and Downes (1996) have suggested that the CO velocity pattern along the arms 
is consistent with gas flow along $x_{1}$-orbits, and that the broader CO 
line widths near the center of the galaxy may be due to the existence of 
the $x_{2}$-orbit. 
Here we show CO (2-1) maps of the central region of NGC~4258, at $3.^{\prime \prime}0\times2.^{\prime \prime}0$ resolution, 
obtained with the Submillimeter Array
\footnote{
The Submillimeter Array is a joint project between the Smithsonian Astrophysical Observatory and the Academia Sinica Institute of Astronomy and Astrophysics, and is funded by the Smithsonian Institution and the Academia Sinica.
}
 (SMA; Ho, Moran and Lo 2004), 
and compared the CO (2-1) maps with the past results of CO (1-0) 
by BIMA (Helfer et al. 2003). 
We evaluate the motion of the gas close the nucleus, and we present models in order to explain the velocity field of the CO 
molecular gas.

\section{Observations and Data Reduction}


The observations were carried out on May 4 2004 with the SMA, 
which consists of eight 6-m antennas. 
All the antennas were used for the observations. 
Projected baseline lengths ranged from 10~m to 128~m, 
which resulted in a synthesized beam size of 
$3.^{\prime \prime}0 \times 2.^{\prime \prime}0$ 
(104~pc $\times$ 69~pc in NGC~4258). 
Zenith opacity was between 0.1 and 0.3 at 225~GHz during the observations, 
and typical double sideband system temperatures 
towards the galaxy were 100--200~K at 1.3~mm. 
The  correlator was configured  
with a frequency resolution of 0.8125~MHz.
We observed J1150+497 and J1308+326 every 30 minutes 
for system gain calibrations. 
Jupiter and Callisto were observed for flux and bandpass calibration, respectively. 

Calibration of raw data was done using the MIR package.  
Imaging and data analysis were performed  
using the MIRAD and NRAO AIPS package. 
Channel maps were made every 10~km~s$^{-1}$ 
with natural weighting of the u,v data.  
CO (2-1) emission was detected in 46 channel maps and imaged 
with the CLEAN method. 
The integrated intensity map and isovelocity map were created 
from the 46 CLEAN channel maps with CO (2-1) emission.  
No continuum emission was detected from data integrated over 
all line-free channels in a frequency range of 1.1~GHz 
down to  a 3~$\sigma$ rms noise level of 8.7 mJy~beam$^{-1}$. 
The FWHM of the primary beam field of view 
with the SMA antenna at 230~GHz is $53^{\prime \prime}$.  
For comparison with CO (1-0), 
we used the BIMA archival channel maps. 
The synthesized beam size of the BIMA maps is 
$6.^{\prime \prime}1 \times 5.^{\prime \prime}4$.




\section{Results}
\subsection{CO (J=2-1) integrated intensity}

CO (2-1) emissions is detected within the velocity range from 200 to 655 km~s$^{-1}$, 
which is consistent with past single-dish observations 
with the IRAM 30-m telescope (Cox and Downes 1996). 
Convolving with the same beam size as the single dish 
($12.^{\prime \prime}5$), 
we estimate that our observations recovered $\sim60\%$ of the single-dish flux. 
The missing flux is due to the lack of short ($<$ 10~m) baselines, 
corresponding to a component bigger than 26 arcsec. 

\begin{figure}
\epsscale{.90}
\plotone{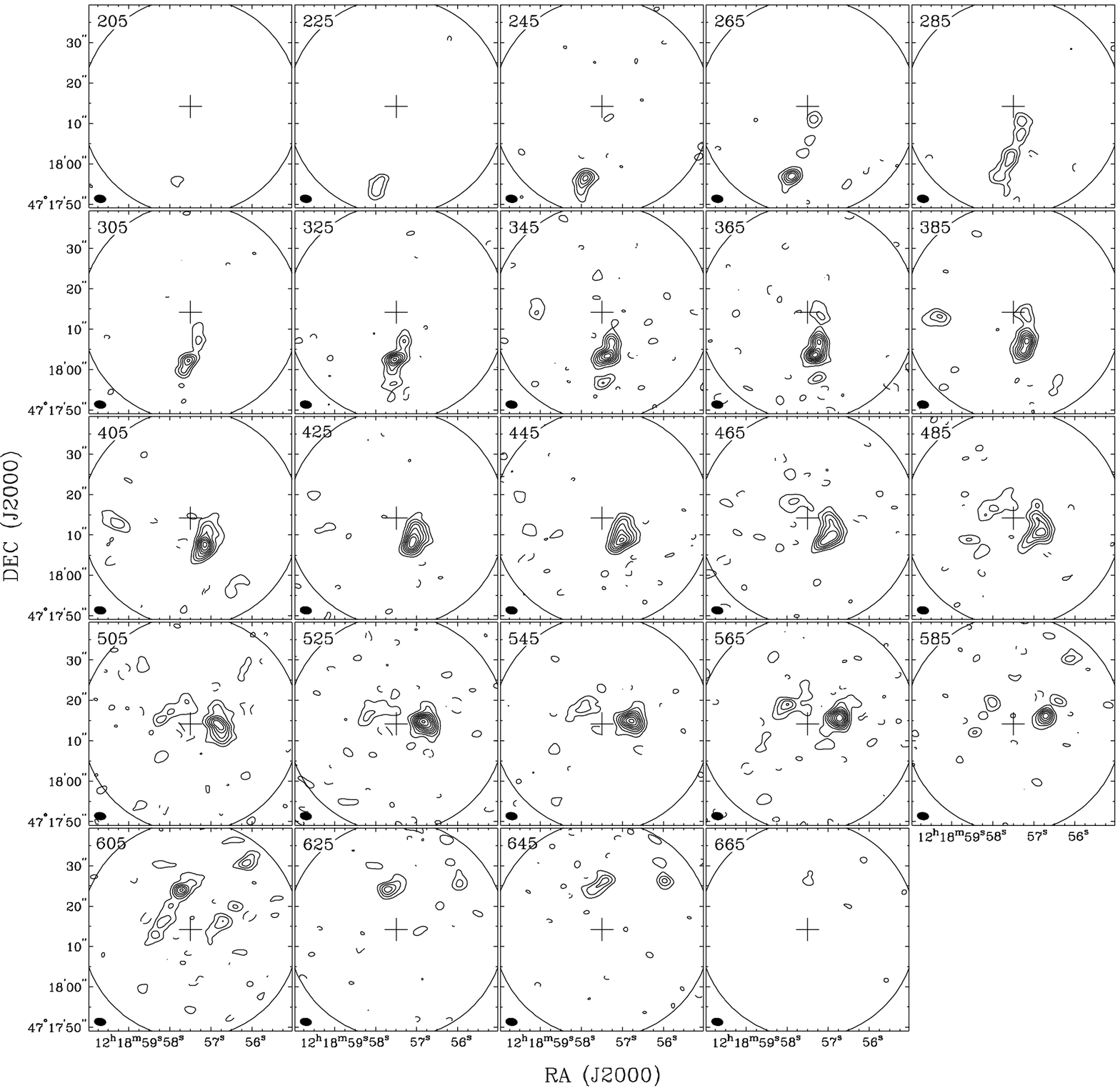}
\caption{Channel maps of CO (2-1) line in NGC~4258 averaging 
every 20~km~s$^{-1}$.  Achieved $\sigma$ level is 0.03~Jy~beam$^{-1}$. 
The contours start at $\pm 3 \sigma$ in steps of $3~\sigma$. 
Their central velocities are shown at the upper left. 
The circle and cross sign indicate the field of view 
($53^{\prime \prime}$) 
and the galactic center (RA$=$12h18m57.5s, DEC$=+$47d18m14s 
by Turner and Ho 1994).  
The synthesized beam size is shown at the lower left. 
Primary beam correction has not been applied to 
the channel maps.
}
\label{channel}
\end{figure}

\begin{figure}
\epsscale{.80}
\plotone{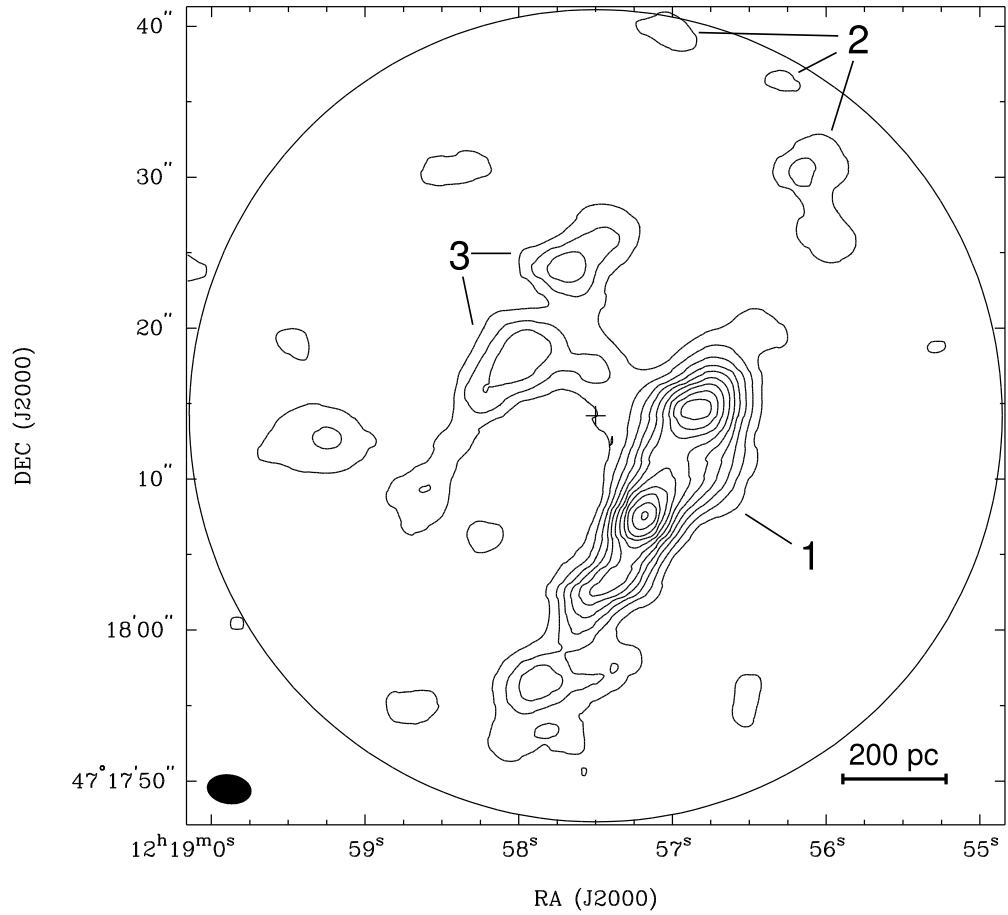}
\caption{Integrated CO (2-1) intensity map in NGC~4258.
Components identification follows the nomenclature 
by Plante et al. (1991). 
The contours start at $\pm 3 \sigma$ in steps of $3~\sigma$,  
where $\sigma=2.1$~Jy~beam$^{-1}$~km~s$^{-1}$. 
The circle and cross sign indicate the field of view 
($53^{\prime \prime}$) 
and the galactic center (RA$=$12h18m57.5s, DEC$=+$47d18m14s 
by Turner and Ho 1994).  
The synthesized beam size is shown at the lower left. 
Linear scale is shown at the lower right. 
Primary beam correction has been applied to the integrated map. 
}
\label{co21int}
\end{figure}

The channel maps of the CO (2-1) line 
with velocity resolution of  20~km~s$^{-1}$ 
are shown in figure~\ref{channel}. 
The integrated intensity map as shown in figure~\ref{co21int}, 
reveals two main elongated structures along P.A. of 152$^{\circ}$ and 
a few faint components.  
These components identified by Plante et al. (1991) are labeled 
from 1 to 3. 
The west elongated structure (Component 1) shows 
several bright peaks, 
which is resolved for first time with the higher angular resolution of the SMA. 
The east elongated structure (Component 3) was 
resolved into two sub components,  
which have been found as a double-peak structure 
by BIMA (Helfer et al. 2003).  
The southern sub component of Component 3 is linked to Component 1 by a faint ridge. 
In previous CO (1-0) maps, Component 2 was an elongated component 
connecting to the north end of Component 1
(Plante et al. 1991, Helfer et al. 2003).  
Our CO (2-1) map shows that 
Component 2 appears only as a few resolved clumps located north 
of Component 1 
possibly due to the missing flux or  
due to lack of CO (2-1) emission. 
Although the extended structures of the center 
found in the IRAM CO (2-1) map (Cox and Downes 1996) 
and in the BIMA CO (1-0) map
are not seen in our SMA map, 
the general agreement of the CO (2-1) emission with the  
CO (1-0) emission is very good. 
Note that there is
no concentrated CO (2-1) emission in the center of the galaxy as in the case of  M51 (Matsushita et al. 2004).

The integrated intensity at the position of the galactic nucleus, 
is at the 3~$\sigma$ level 
6.3~Jy~beam$^{-1}$~km~s$^{-1}$ ($=$24.6~K~km~s$^{-1}$). 
Assuming the ratio of integrated intensity CO(2-1)/CO(1-0) $=1$,
we can estimate the column density of molecular hydrogen 
$N_{\rm H_{2}}=7.5\times10^{21}$~cm$^{-2}$ in the center 
using a Galactic CO-H$_2$ conversion factor of 
$3.0 \times 10^{20}$ cm$^{-2}$~(K~km~s$^{-1}$)$^{-1}$ 
(Solomon et al. 1987).  
Equivalent $N_{\rm H}$ would be $1.5\times10^{22}$~cm$^{-2}$. 
The value is lower than the estimate of 
$N_{\rm H}=5.9$-$13.6\times10^{22}$~cm$^{-2}$ 
as suggested by the modeling of X-ray spectra (Fruscione et al. 2005). 
However the estimates of $N_{\rm H}$ from the X-ray data refer 
to a much smaller sizescale associated with the accretion disk around 
the central AGN. 
Total mass of the molecular gas in the nuclear region within the 
central 100~pc is estimated to be $<4.5\times10^6~M_{\odot}$.

\subsection{CO (J=2-1) isovelocity field}

\begin{figure}
\epsscale{.80}
\plotone{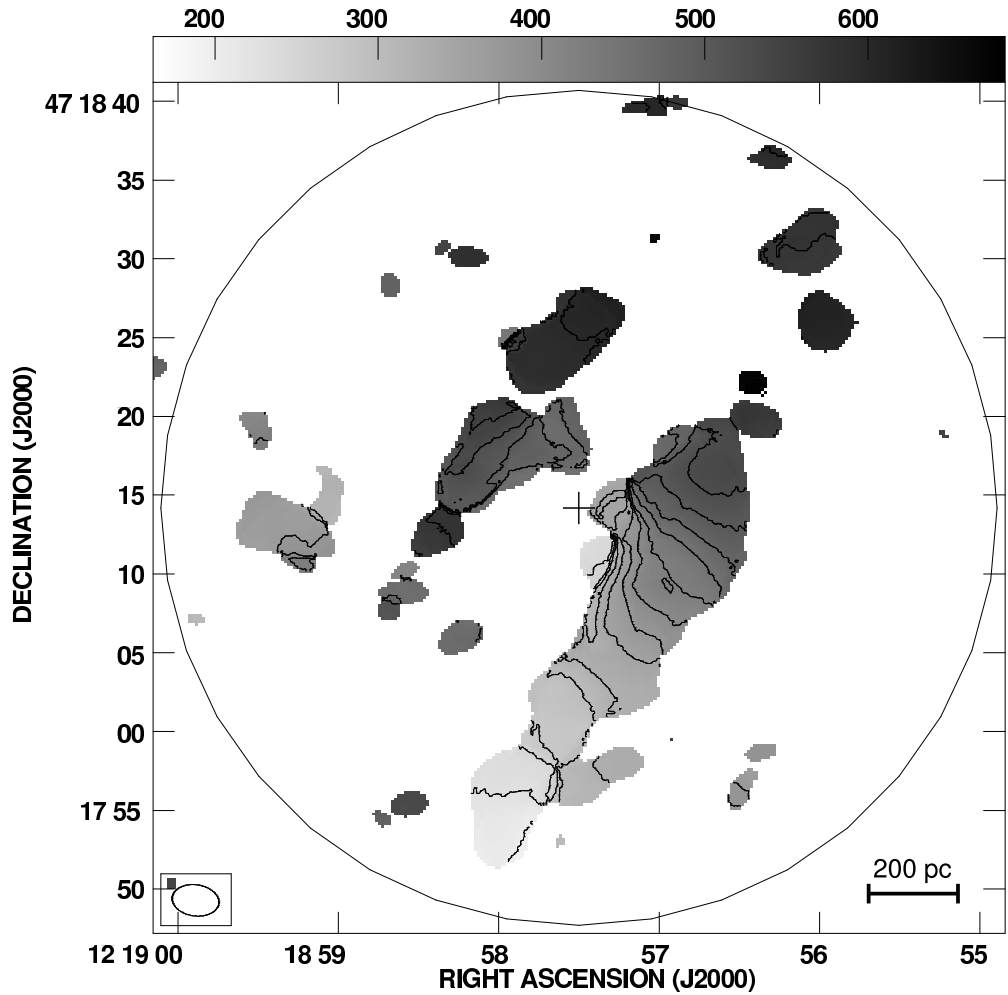}
\caption{Isovelocity CO (2-1) map in NGC~4258. 
Contour levels are 25~km~s$^{-1}$ interval from 200~km~s$^{-1}$.  
The circle and cross sign indicate the field of view 
($53^{\prime \prime}$) 
and the galactic center (RA$=$12h18m57.5s, DEC$=+$47d18m14s 
by Turner and Ho 1994).  
The synthesized beam size is shown at the lower left. 
Linear scale is shown at the lower right. 
}
\label{co21vel}
\end{figure}

\begin{figure}
\epsscale{.80}
\plotone{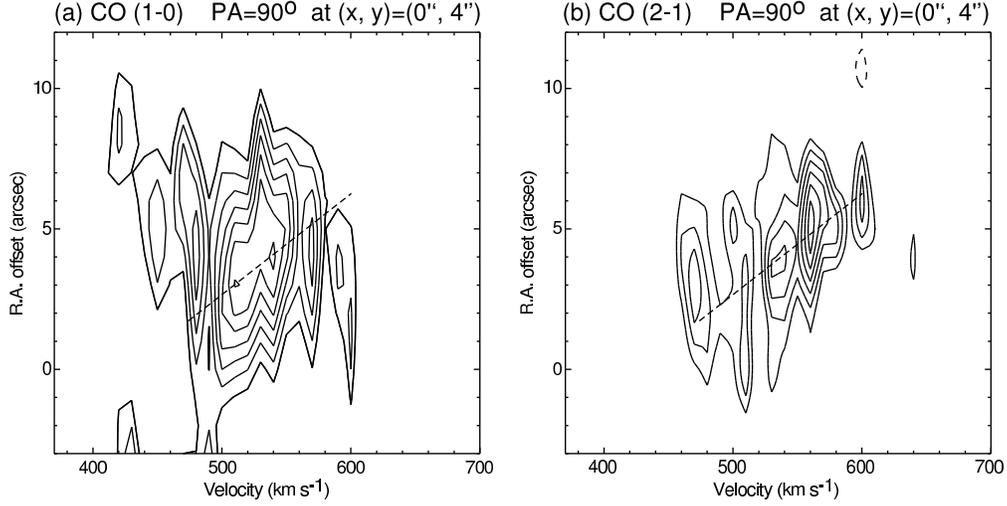}
\caption{
Position-Velocity diagram along P.A.$=$90$^{\circ}$ 
on the southern part of Component 3 
in (a) CO (1-0)  and (b) CO (2-1). 
We cut the line for the positional-velocity diagram 
to run through a point with position of (0, 4) arcsec. 
The dashed line in (a) connects the two peaks at 510 and 540~km~s$^{-1}$. 
The exactly same dashed line as (a) is superposed on (b), and the 
dashed line agrees well with the velocity gradient in (b). 
The velocity resolution is 10~km~s$^{-1}$.  
Contour levels are (a) $-3, 3, 4, 5, 6, 7$ and $8 \sigma$, 
where $1\sigma=42$~mJy~beam$^{-1}$, and 
(b) $-3, 3, 4, 5, 6$ and  $7 \sigma$, 
where $1\sigma=40$~mJy~beam$^{-1}$.  
}
\label{pvco1021}
\end{figure}

Figure~\ref{co21vel} shows the isovelocity contour map of the CO (2-1) 
emission in NGC~4258. 
Contours of the velocity field in Component 1 are almost perpendicular 
to the optical major axis (P.A.$=150^{\circ}$) of the galaxy. 
Component 1 is extended over 1.1~kpc with a 
velocity range from 200 to 550~km~s$^{-1}$ along the major axis. 
The velocity contours for Component 1 are more closely spaced and 
rotated by $\sim30^{\circ}$ as we get closer to the center, 
which agrees with the velocity field of CO (1-0) which Plante et al. (1991) have shown.
The velocity field on Component 2 extends from 550 to 650~km~s$^{-1}$
and the velocity contours within Component 2 are almost perpendicular to 
the direction of P.A. of $\sim20^{\circ}$.  
The velocity pattern on Component 2 also agree well with those of CO (1-0). 
Within Component 3, we see two different velocity gradients. 
One velocity gradient is roughly seen 
along the major axis of Component 3, 
and similar to the CO (1-0) map. 
The other velocity gradient  is almost 
along the east-west direction on the southern sub component of Component 3, 
which cannot be explained by a simple galactic rotation. 
The position-velocity diagrams on Component 3 
along the P.A.$=90^{\circ}$
in CO (1-0) and CO (2-1) is shown in figure~\ref{pvco1021}.  
In CO (2-1), the velocity gradient is apparently found along 
the P.A. of 90$^{\circ}$ 
with a velocity range from 470 to 600~km~s$^{-1}$.  
Velocity gradient in CO (1-0) reveals two different velocity gradients; 
the first gradient is shown from 400 to 600~km~s$^{-1}$,
which is due to the velocity gradient along the major axis of Component 3. 
The second gradient is formed by  two peaks 
at 510 and 540~km~s$^{-1}$, 
as shown by the dashed line in figure~\ref{pvco1021}(a), 
which is similar to the velocity gradient found in CO (2-1).

\subsection{Residual velocity field after subtracting the rotation model}

\begin{figure}
\epsscale{.80}
\plotone{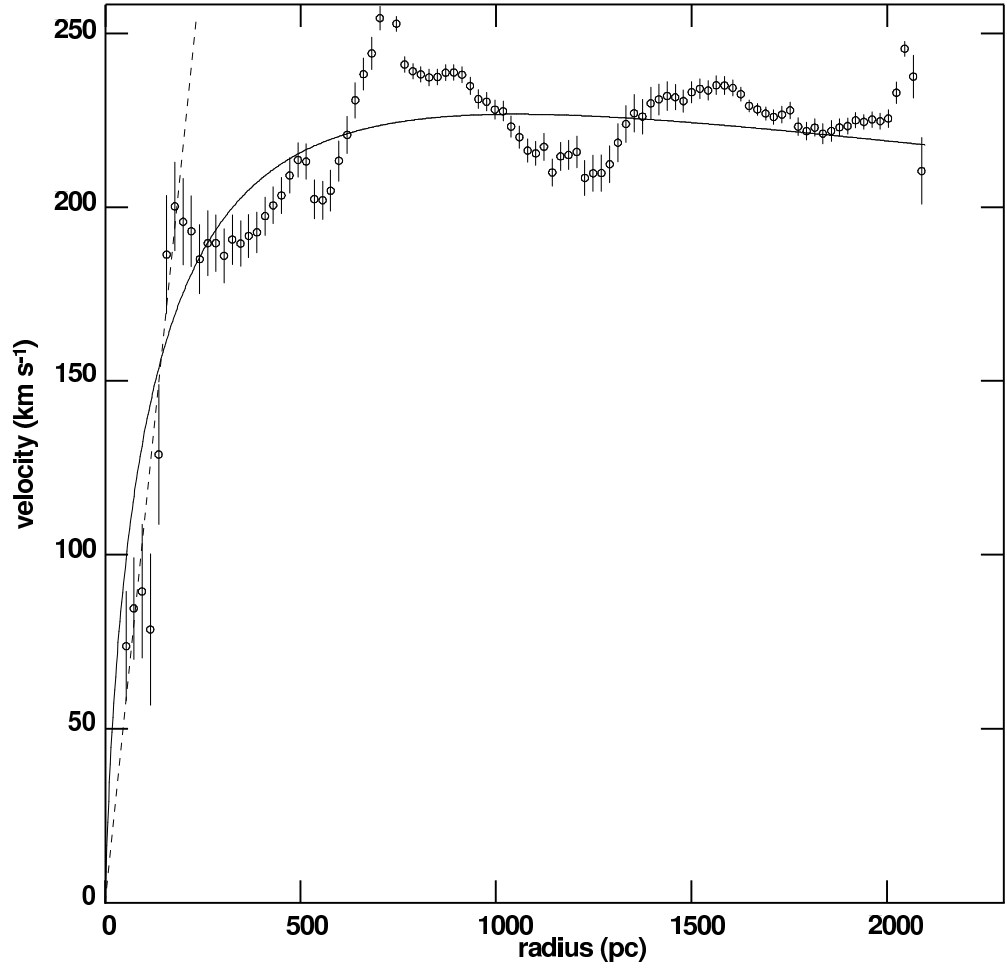}
\caption{
Brandt rotation curve for NGC~4258.  
Each data points are 
the average values of velocity and the standard deviation at any radius. 
Dashed line indicates the solid body rotation within 200~pc. 
}
\label{rotationcurve}
\end{figure}

In order to see the unusual velocity gradient clearly, 
we analyzed the velocity field assuming circular motions 
to subtract a rotation curve. 
We performed a least square fit by applying the AIPS 
GAL task (van Moorsel and Wells 1985). 
Because the velocity fields of CO (1-0) and CO (2-1) are so similar to each other,  
while the field of view of the CO (1-0) map is 
larger and the CO (1-0) emission is more extended,  
we modeled the overall velocity field by fitting the BIMA CO (1-0) 
map with a Brandt rotation curve (Brandt 1960),  
\begin{equation}
\frac{V}{V_{max}}=\frac{R/R_{max}}{(1/3+2/3 (R/R_{max})^{n})^{(3/2n)}}, 
\end{equation}
where $n$ is a measure of the steepness of the rotation curve and 
$R_{max}$ is the radius 
at which the maximum rotation velocity $V_{max}$ occurs.
The fitting was done with a range of radius from 0 to 60~arcsec (0 to 2~kpc) 
in the center of NGC~4258. 
Figure~\ref{rotationcurve} shows the velocity along the radius and 
the resulting rotation curve.  
The fit with GAL also gives parameters of center position, 
position angle, inclination and systemic velocity (table~\ref{parameter}).
The agreement with previous estimates is excellent.   
In figure~\ref{rotationcurve}, the rotation curve drops rapidly at 200~pc. 
Using the resulting parameters of the center position, the position angle and the inclination from the 
Brandt rotation in table~\ref{parameter},   
the velocities within 200~pc is also fitted by a solid body rotation. 
From the Brandt rotation, the dynamical masses 
within the radii of 200~pc and a 320~pc are obtained 
as $1.3\times10^{9}~M_{\odot}$ and $2.9\times10^{9}~M_{\odot}$, 
respectively. 

From the solid body rotation, the dynamical mass 
within the 200~pc radius is 
estimated as $2.0\times10^{9}~M_{\odot}$. 
These estimates are consistent with the dynamical mass of $3\times10^{9}~M_{\odot}$
estimated to be within a 320~pc radius 
from previous observation of CO (1-0) in NGC~4258 (Sofue et al. 1989).  
They are also similar to the dynamical mass measured with CO (1-0) 
within a 200~pc radius for the Virgo spiral galaxies 
(0.1--3$\times10^{9}M_{\odot}$; Sofue et al. 2003).

\begin{figure}
\epsscale{0.7}
\plotone{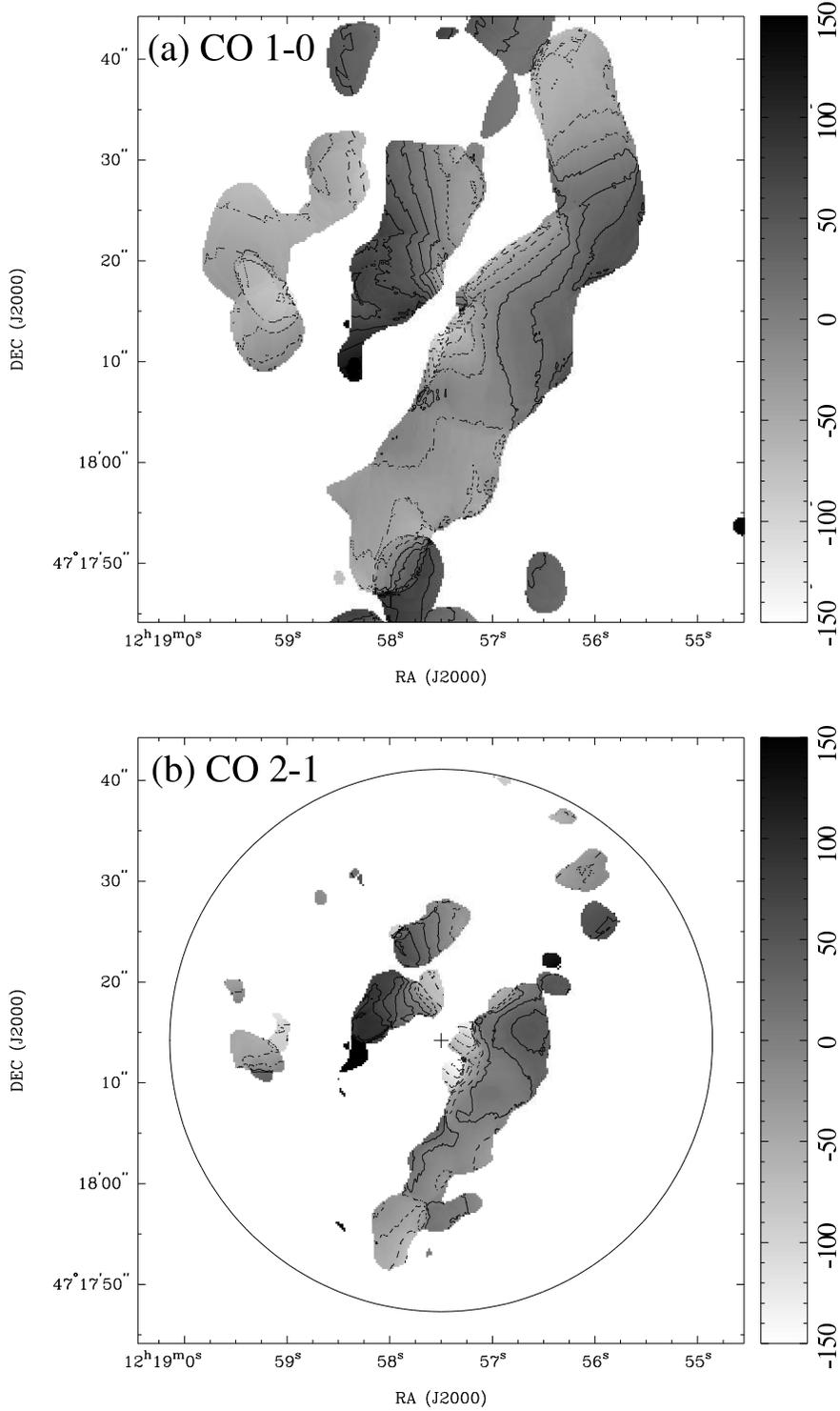}
\caption{
Residual velocity field after subtracting the Brandt rotation 
model in (a) CO (1-0)  and (b) CO (2-1). 
Contour levels are 25~km~s$^{-1}$ interval from $-150$~km~s$^{-1}$ 
to $150$~km~s$^{-1}$.  
The circle and the cross sign in (b) indicate the field of view 
($53^{\prime \prime}$) 
and the galactic center, which is same as them seen in figure~\ref{co21int} 
and figure~\ref{co21vel}. 
Beam sizes are;  
(a) $6.^{\prime \prime}1\times5.^{\prime \prime}4$ and P.A. of 4.1$^{\circ}$,  
and (b) $3.^{\prime \prime}0 \times 2.^{\prime \prime}0$ and P.A. of 82$^{\circ}$.  
The velocity field of the Component 3 deviates from that 
of the galactic rotation.}
\label{subtract}
\end{figure}

Residual velocity field maps are obtained by subtracting the Brandt rotation model 
from the CO (1-0) and CO (2-1) isovelocity map (figure~\ref{subtract}). 
The residual velocity field is within a range of $\pm50$~km~s$^{-1}$ 
on Component 1 and 2.  
Within the southern sub component of Component 3, 
a large velocity  offset of $\pm 150$~km~s$^{-1}$  remains. 
We clearly find a residual velocity gradient 
along the east-west direction 
which spans $\sim$6 arcsec (200~pc)
in the southern sub component of Component 3 (figure~\ref{subtract}b).

We can compare the residual velocity field in CO (2-1) 
with a residual velocity map of H$\alpha$ 
after subtracting a rotation model in the center, 
which is similar to the Brandt rotation. 
The H$\alpha$ results have shown that the velocity 
in the anomalous arms is considerably 
offset from the basic rotation (van der Kruit 1974),  
with in the northern arm blueshifted 
by  $\sim-100$ to $-150$~km~s$^{-1}$ 
in the northern part of the galactic center. 
However, the large velocity offset of H$\alpha$ is seen 
$\sim30$ arcseconds (1~kpc) shifted to north-west from the center.  
The H$\alpha$ residual velocity map shows no significant velocity offset 
($<$ 50~km~s$^{-1}$) at the region where Component 3 lies, 
 $\sim$200~pc shifted to north-east from the center.  
It is possible that  the velocity offset in H$\alpha$ further out 
is due to a jet from the 
central engine, 
while the velocity offset in CO (2-1) near the center could be a reflection of 
another motion in the circumnuclear region.

\subsection{The CO(2-1)/CO(1-0) line ratio map}

\begin{figure}
\epsscale{1.00}
\plotone{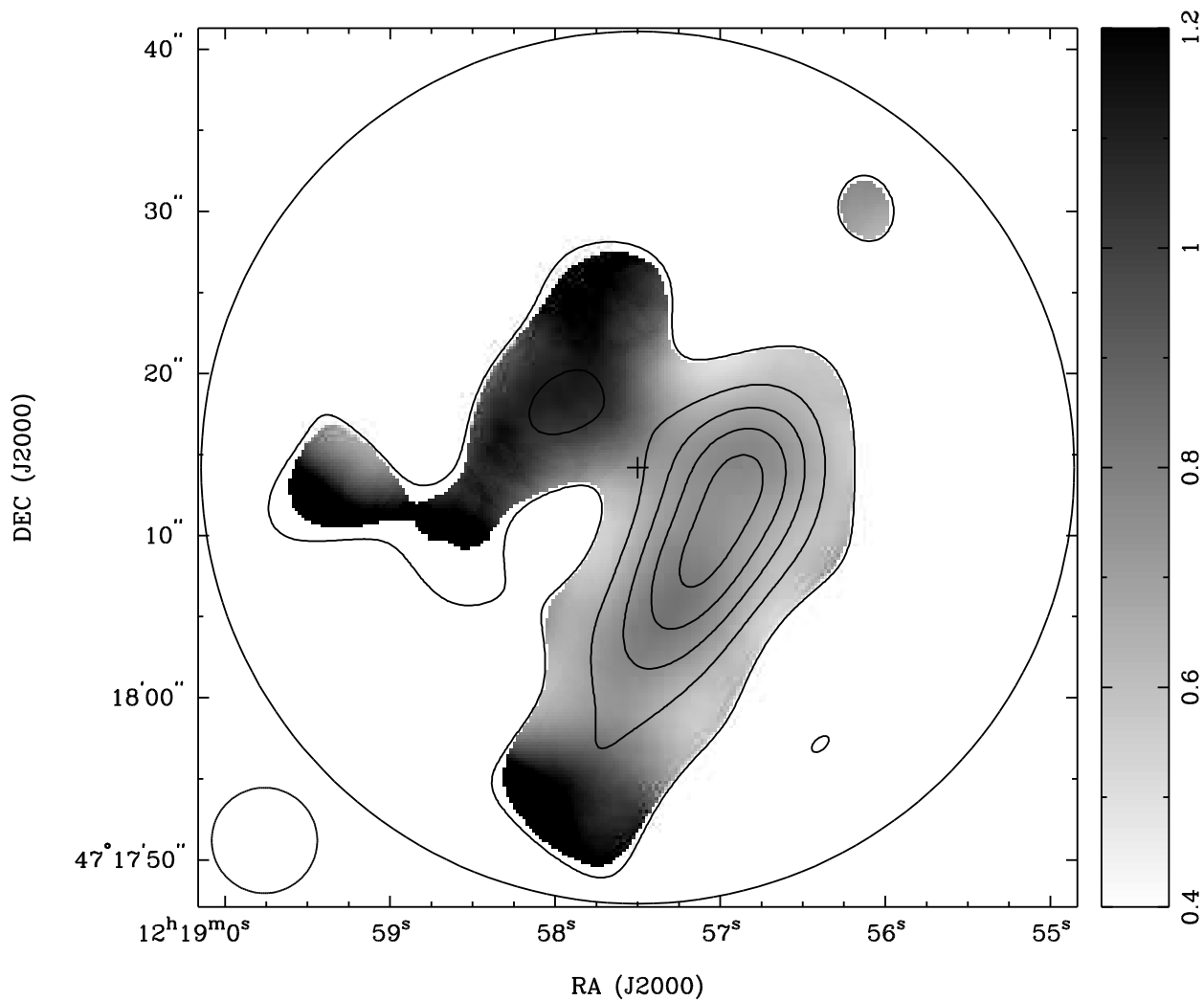}
\caption{Line ratio map of CO(2-1)/CO(1-0) 
from the SMA CO (2-1) map and the BIMA CO (1-0) map with the 
SMA uv coverage. 
 The two maps has same uv sampling and were smoothed 
 by a $6.^{\prime \prime}5 \times 6.^{\prime \prime}5$ beam. 
 The smoothed beam size is shown at the lower left.
Grey scale shows that line ratio of CO(2-1)/CO(1-0). 
Contours indicate the smoothed CO (2-1) with levels of  
3, 6, 9, 12 and 15$\sigma$ before the primary beam correction. 
The big circle and cross sign are the field of view 
(53$^{\prime \prime}$) and the galactic center. 
}
\label{lineratio}
\end{figure}

The ratio of the CO brightness temperatures allows us to
estimate the physical state of the CO gas. 
In order to obtain a CO (1-0) intensity map  
with the same uv sampling as our SMA CO (2-1) data, 
we used the CO (1-0) BIMA channel maps as a model. 
The CO (1-0) BIMA map includes single-dish data and 
it recovers extended structures. 
We performed FFT and sampled it with our SMA uv coverage.
Next we imaged channel by channel and 
created a new integrated map from these sampled visibilities.  
Then we deconvolved it in the usual way. 
As the CO (1-0) data are missing the long uv spacings, 
both the CO (1-0) simulated image and the SMA CO (2-1) image 
were finally smoothed to the same resolution of 
$6.^{\prime \prime}5\times6.^{\prime \prime}5$.
We found that about 30 to 40 $\%$ of the flux was lost 
from the BIMA CO (1-0) model which includes single dish data, 
due to the lack of short baselines ($<10$ m) for the SMA uv coverage. 
This is similar to the percentage of missing flux estimated in the SMA CO (2-1) map. 
We combined this simulated CO (1-0) map with our CO (2-1) map to produce a CO(2-1)/CO(1-0) brightness temperature ratio map 
which is shown in figure~\ref{lineratio}. 
The primary beam attenuation has been corrected in 
the CO (2-1) and CO (1-0) maps.  
Finally, we clipped out the pixels 
where the signal to noise ratio of CO(2-1)/CO(1-0) was lower than 3.
The line ratio map after the primary beam correction (figure~\ref{lineratio}) reveals that the 
average of the line ratio on pixels in Component 3 is higher ($1.1\pm0.2$) as compared to the ratio 
in Component 1 ($0.7\pm0.1$).  
The value of line ratio in Component 2 has a relatively large error ($0.7\pm0.3$), 
as the Component 2 lies near the edge of the primary beam. 
These values agree with the past surveys of CO line ratios 
in various galaxies (e.g. Braine et al. 1993, Aalto et al. 1995). 
A gradiation of gas excitation consisting of warm gas in the 
central regions and cold gas in the outer regions can explain  
the difference of line ratio among the components.  
Such a difference of the line ratio between the inner nuclear region and the outer parts of a galaxy, has already been found;  
Papadopoulos and Seaquist (1998) showed the correlation 
between the CO(2-1)/CO(1-0) line ratio and the ratio 
of the source size to the beam size.  
The value of the line ratio in Component 3 is close to    
the average value of 0.9 
found for the nuclear regions ($<$ 1~kpc) 
of Seyfert and/or starburst galaxies 
(Papadopoulos and Seaquist 1998).  
We estimated the CO molecular gas characteristics using a LVG analysis. 
For a kinetic temperature of 30~K, 
the  molecular hydrogen density of the gas on Component 1 is 
of the order of 100~cm$^{-3}$.  
On the other hand, the density of the gas 
on Component 3 could be of the order of 1000~cm$^{-3}$ 
or more.

\section{Modeling of the gas kinematics in the central region}

The significant differences of the CO(2-1)/CO(1-0) ratio and 
the residual velocity offsets within Component 1 and 3 may imply 
that 
the physical conditions of these two components are not the same.  
One possible reason for the difference may be   
that Component 3 is located near the center as compared  
with Component 1.  
Gas heating by the central engine 
or perturbation by nuclear activities could exist. 
Such effects would be less significant for Component 1, if it is 
located further out from the nucleus. 
We cannot say too much about Component 2 as the value 
of the CO(2-1)/CO(1-0) ratio is not well determined. 

The velocity field of the central region in NGC~4258 has several 
observed characteristics; 
(1) There is a main velocity gradient along the major axis on Component 1 and 3, 
(2) The velocity contours are more tightly spaced and tilted with 
respected to background motions as we approach the nucleus 
within Component 1, and  
(3) There is a similar velocity gradient within Component 3 
along the east-west direction. 
The bulk of the gas within Component 1 could be further out 
in the galactic disk, because the velocity pattern on Component 1 
is consistent with the galactic rotation.   
On the other hand, the velocity offset of Component 3 from the 
galactic rotation is larger and it can be due to some nuclear activities 
close to the center.

\begin{figure}
\epsscale{1.00}
\plotone{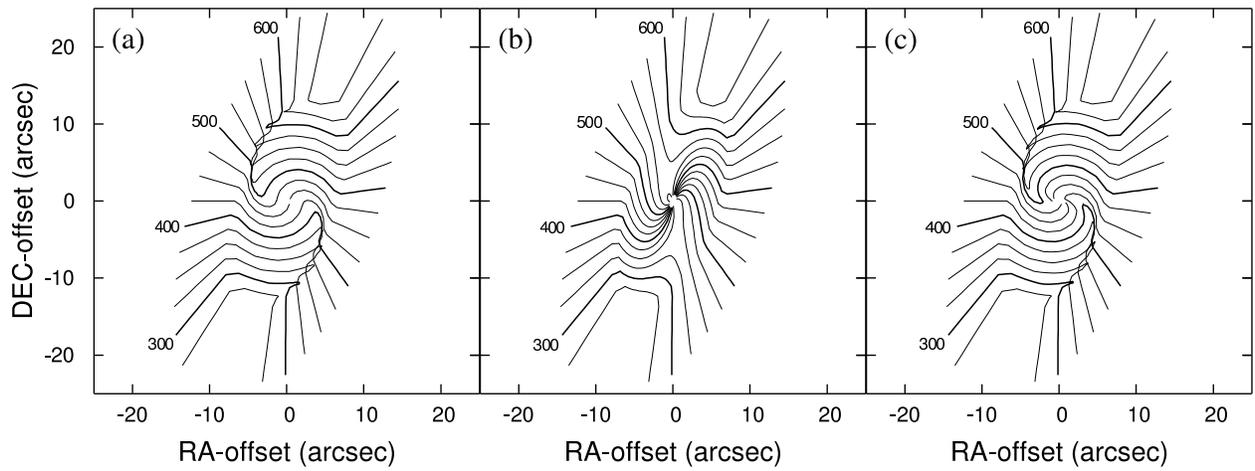}
\caption{
Velocity fields of the rotation models featuring (a) warped disk, (b) expansion disk  
and (c)  warped expansion disk. 
Velocity contour levels are 25~km~s$^{-1}$ interval from 200~km~s$^{-1}$. 
We adopt the circular solid-body rotation with P.A. of $-20^{\circ}$ 
for all models. 
The warped disk is tilted from P.A. of $90^{\circ}$ to $-20^{\circ}$. 
}
\label{model}
\end{figure}

To account for the characters of the velocity field, 
we propose models of expanding and warped disks.  
As we have shown in table~\ref{parameter},  we used the  
galactic rotation with the P.A. of 160$^{\circ}$, 
the inclination of 65.6$^{\circ}$ and 
the $V_{\rm sys}$ of 456~km~s$^{-1}$. 
Figure~\ref{model} shows the isovelocity contour maps of the 
warped disk model, the expansion disk model, and the combined model with warped disk and the expansion. 
As we discuss below, 
the warped disk model and the combined model of 
expansion and warped disk in figure~\ref{model}(a) and (c) 
are preferable to explain the velocity field.

Because the trend of gradient is same,  
we can conjecture that the east-west velocity gradient which extends $\sim$200~pc is related to 
 the sub-parsec disk surrounding the central engine  
(Miyoshi et al. 1995, Herrenstein et al. 1999), 
although the east-west velocity gradient looks a solid body rotation with the angular resolution of 3$^{\prime\prime}$ 
and does not seem to be the declining Keplerian tails of the sub-pc molecular disk.  
The solid rotation velocity estimated from the position-velocity diagram of figure~\ref{pvco1021} 
at a radius of 3$^{\prime\prime}$ 
would be 74~km~s$^{-1}$, and exceeds the velocity estimated from the 
Keplerian rotation at the same radius (39~km~s$^{-1}$). 
If we assume that gas within Component 3 is closer to the center 
and that the east-west velocity gradient is due to rotation, 
the enclosed binding mass is estimated to be 
$2 \times 10^{9}$~$M_{\odot}$ within 200~pc. 
This mass is similar to the enclosed mass obtained from the fitted  Brandt rotation curve ($1.3\times10^{9}~M_{\odot}$ within 200~pc). 
Therefore, the velocity gradient along the east-west direction on Component 3 
could be due to a circular motion of the gas 
in a different orbital plane from the galactic plane, 
but which aligns with the inner sub-parsec disk. 
However, the scenario of the two different orbital planes alone 
does not explain the titled velocity contours. 
If the major axis of the orbit warps 
from a P.A. of 90$^{\circ}$ to a P.A. of $-20^{\circ}$, 
the velocity contours close to the center would be more tightly spaced  and shifted. 
This can explain the velocity contours of the inner edge of the Component 1 (figure~\ref{model}a), 
The velocity gradient along the major axis of Component 1 and 3 
would be due to the galactic rotation. 
The differences in the magnitude of the gradients could be due to the fact that these components are not 
at the same radial distance along the entire structure. 
If the gas on Component 3 is closer to the center as the line ratio suggests, 
the two different 
velocity gradients along the major axis and the east-west direction 
might be explained.

An alternative scenario to explain the velocity field  
in the center is a uniform expansion from the center of the galaxy 
as Plante et al. (1991) has  proposed. 
The model can explain the angular offset of the velocity contours 
for Component 1 by $\sim30^{\circ}$ and 
the velocity gradient along the east-west direction 
seen for Component 3. 
Component 1 and 3 are thought to be located front and back from the center, respectively. 
In the case of uniform expansion, 
Component 1 and 3 are on opposite sides from each other. 
The expansion pushes gas in front and back to 
blueshifted and redshifted velocity from the center, respectively. 
The east-west velocity gradient seen in Component 3 
can be due to projection 
along the line of sight of radial expasion motions.  
It is reasonable that we see the velocity gradient 
mostly within Component 3 
because Component 1 may be further from the center. 
The scenario of expansion can also support the idea 
that the gas closed to the center is more excited and therefore
has a higher value of the line ratio CO(2-1)/CO(1-0). 
One problem is that the observed velocity fields show 
the most redshifted velocity on the eastern edge of Component 3. 
The velocity pattern of an expanding disk would show 
the most redshifted gas closer to the center (figure~\ref{model}b). 
Thus, expansion alone does not account for the observed 
velocity field.

In NGC~4258, the presence of a bar has been proposed, 
although the proposed P.A. of the bar varies  
from $-35^{\circ}$ to $17^{\circ}$  
(e.g. van Albada and Shane 1975; van Albada 1980; Cox and Downes 1996; 
Cecil et~al. 2000). 
Here we discuss the possibility that Component 1 and 3 are 
molecular gas produced by bar driven shocks,  
and that they flow in and around the bars 
into the periodic orbit perpendicular to the arms 
(e.g.  Athanassoula 1992; Downes et al. 1996). 
Velocity gradients seen in Component 1 and 3 along the major axis 
are roughly consistent with the $x_{1}$ -orbit, 
as Cox and Downes (1996) have mentioned already. 
However, the bar model does not describe  
the velocity gradient along the east-west direction 
seen in Component 3. 
The velocity pattern of the periodic $x_{2}$ -orbit would show 
the most redshifted 
velocity in northern side of the major axis of the galaxy, and 
values of velocity would decrease continuously 
as it goes to west along the east-west direction.  
This is opposite to the velocity gradient in our isovelocity map. 
Therefore, we conclude that the CO gas does not trace the bar structure.  

Here we present a combined model of the expanding warped disk.  
The more tightly spaced and tilted velocity contours  
at the inner edge of Component 1, and 
the east-west velocity gradient seen in Component 3 are,  
clearly seen in the expanding warped disk model. 
The main velocity gradient along the major axis of Component 1 
and 3 could be due to galactic rotation. 
If the gas on Component 3 lies at an inner orbit, 
the velocity gradient along its major axis may be steeper as 
compared to Component 1.  
Thus, the expanding warped disk can more easily model all the 
observed characteristics of the observed velocity field 
(figure~\ref{model}c).

\section{Conclusions}

We conducted SMA observations of the CO (2-1) emission 
toward the center of NGC~4258 with a high spatial resolution of $\sim100$~pc. 
The CO (2-1) integrated intensity map is similar to that of CO (1-0) and 
does not show a concentrated component in its center. 
The line ratio map of CO(2-1)/CO(1-0) suggests that 
the gas within Component 3 is warmer and denser as compared with that of Component 1. 
This is consistent with Component 3 being 
close to the center of the galaxy 
and possibly perturbed by the central nuclear activities.   
CO (2-1) and CO (1-0) velocity field maps correlate well with 
each other, 
and the major feature of the observed velocity field can be explained  by the rotation curve. 
However, the velocities in several small regions, 
including Component 3, deviate from this rotation curve. 
In order to account for the velocity field, 
we propose kinematical models  
for the rotating disk with expanding and warped features.

\acknowledgments
We are grateful to all the staffs of SMA for their support during our observations. 
We also express our appreciation to BIMA for the archival BIMA SONG images of NGC~4258. 
We appreciate to Drs. K. Kohno,  
S. Kameno and Y. C. Minh for their helpful comments. 
S. S-S would like to thank Drs. N. Hirano and S. Takakuwa for their 
thoughtful help for analysis of the data. 




Facility: \facility{SMA}

\clearpage



\clearpage

\begin{table}
\begin{center}
\caption{Parameters of fitting \label{parameter}}
\begin{tabular}{lccccc}
\tableline \tableline
  Parameters  &  Our results   & \multicolumn{4}{c}{Previous results}   \\
  \cline{3-6}
                     &	CO (1-0)	&  Radio continuum$^{a}$   &  Water maser &  HI$^{d}$  &  H$\alpha^{e}$  \\
\tableline 
RA of center (J2000)    & 12:18:57.52       & 12:18:57.50 & 12:18:57.5046$^{b}$ &         &  \\
DEC of center (J2000) & $+$47:18:15.98 & $+$47:18:14 & $+$47:18:14.303$^{b}$ &         &  \\
Position angle  (deg.)   &  160                   &                    & & 150  & 146  \\
Inclination (deg.)          &  65.6                  &                     & &  72   &  64  \\
V$_{\rm sys}$ in Radio LSR (km~s$^{-1}$) & 456   &  & 473.5$^{c}$  & 450  &  467  \\
\tableline
\end{tabular}


\tablenotetext{a}{Brightest position at 6~cm radio continuum (Turner and Ho 1994).  }
\tablenotetext{b}{Position of the systemic maser at 510 km~s$^{-1}$ (Herrnstein et al. 2005).}
\tablenotetext{c}{The systemic velocity of the maser disk (Herrnstein et al. 2005).}
\tablenotetext{d}{Based on HI emission (van Albada 1980) }
\tablenotetext{e}{Based on H$\alpha$ (van der Kruit 1974)}
\end{center}
\end{table}






\end{document}